\documentclass{aa}

\usepackage{graphicx}
\usepackage{txfonts}

\usepackage{natbib}
\bibpunct{(}{)}{;}{a}{}{,}

\begin{document}

\title{Multiplicity of young stars in and around R Corona Australis%
\thanks{Based on observations obtained at the European South\-ern
Observatory, La Silla, proposal numbers 55.E-0968, 65.I-0086, and
67.C-0213}}

\author{R.\ K\"ohler\inst{1,2}
\and
R.\ Neuh\"auser\inst{3}
\and
S.\ Kr\"amer\inst{3} 
\and
Ch.\ Leinert\inst{2}
\and
T.\ Ott\inst{4}
\and
A.\ Eckart\inst{5}
}
\institute{%
	ZAH Landessternwarte, K\"onigstuhl,
	69117 Heidelberg, Germany, \email{r.koehler@lsw.uni-heidelberg.de}
\and
	Max-Planck-Institut f\"ur Astronomie, K\"onigstuhl 17,
	69117 Heidelberg, Germany
\and
Astrophysikalisches Institut und Universit\"ats-Sternwarte, Friedrich-Schiller-Universit\"at
Jena, Schillerg\"asschen 2-3, 07745 Jena, Germany
\and
MPI f\"ur extraterrestrische Physik, Postfach 1312, D-85741 Garching, Germany
\and
I.~Physikalisches Institut, Universit\"at zu K\"oln,
	Z\"ulpicher Stra{\ss}e 77, 50937 K\"oln, Germany
}

\date{Received April 3., 2008; accepted June 26., 2008}


\abstract
{In star-forming regions like Taurus-Auriga, it has been found that
  most young stars are born as multiples, which theories for star
  formation should definitely take into account. The R CrA
  star-forming region has a small dark cloud with quite a number of
  protostars, T~Tauri stars, and some Herbig Ae/Be stars, plus a
  number of weak-line T Tauri stars around the cloud found by ROSAT
  follow-up observations.}
{We would like to detect multiples among the young stars in and around
  the R CrA cloud in order to investigate multiplicity in this
  region.}
{We performed interferometric and imaging observations with the
  speckle camera SHARP~I at the ESO 3.5\,m NTT and adaptive optics
  observation with ADONIS at the ESO 3.6\,m telescope, all in the
  near-infrared bands JHK obtained in the years 1995, 2000, and 2001.}
{We found 13 new binaries among the young stars in CrA between
  0.13~arcsec (the diffraction limit) and 6~arcsec (set as an upper
  separation limit to avoid contamination by chance alignments).
  While most multiples in CrA are binaries, there are also one
  quadruple (TY CrA), and one triple (HR 7170) which may form a
  quintuple together with the binary HR 7169.  One of the newly
  detected companions with a large magnitude difference found near the
  M3-5 type T~Tauri star [MR 81] H$\alpha$ 17 could be a brown dwarf
  or an infrared companion with an edge-on disk.  Among seven Herbig
  Ae/Be stars in CrA, six are multiple.}
{The multiplicity frequency in CrA is as high as in similar star
  forming regions.  By comparing with the period distribution of
  main-sequence stars and extrapolating to separations not probed in
  this survey, we conclude that the companion-star frequency is
  $(95\pm23)\,\%$; i.e.\ the average number of companions per primary
  is 0.95.}


\maketitle


\section{Introduction}

The star-forming region {\em Corona Australis} (southern crown),
abbreviated {\em CrA} or {\em R CrA}, is today known as one of the
nearest regions of ongoing and/or recent intermediate- and low-mass
star formation. The dark cloud around R CrA has an extinction of up to
$A_{V} \sim 45$ mag.  The age ranges between $\le 1$ Myrs for the
protostars to $9 \pm 4$ Myrs for some T Tauri stars \citep{james2006}.
See \citet{NeuhReview2008} for a recent review.

\citet{cas98} determined the distance towards the eclipsing
double-lined spectroscopic binary (SB) TY CrA to be $129 \pm 11$~pc
from their orbit solution.  Because this is consistent with other
earlier estimates \citep[see discussion in][]{NeuhReview2008} and, at
the same time, the best current estimate, we use $\sim 130$ pc as the
distance for the young stars in CrA.

\citet{NeuhReview2008} give a list of all known optically visible
members compiled from various sources. Only a few multiples have been
published among the CrA members so far; see Table~1 for a listing.
Hence, while six out of seven Herbig Ae/Be stars are known or
suspected to be multiple, only five T~Tauri stars (TTS) were known to
be multiple, a low number compared to other star-forming regions.  In
addition, the brown dwarf member Denis1859 is a known binary
\citep{bouy2004}.

\begin{table*}[ht]
\caption[]{Previously known multiples among optically visible young stars in and around CrA (c,d)}
\label{PrevBin}
\tabcolsep.8em
\begin{center}
\begin{tabular}{lccccccc}

\noalign{\hrule\vskip1.4pt\hrule\vskip1pt}

Designation & Separation & Position         & \multicolumn{3}{c}{Brightness ratio (a)} & Epoch & Reference\\
            & [arc sec]  & Angle [$^\circ$] & J & H & K                         & day/month/year & \\ \hline

S CrA	& $1$		  & $135$	&&&&04/07/1942&(1)\\
	& $1$		  & $134$	&&&&& (15) \\
	& $1.41\pm0.06$   & $157\pm2$	&&& $0.3\pm0.2$ (b)&12/04/1995&(2)\\ 
	& $1.37\pm0.02$   & $147\pm2$	&&&&1981.25&(3)\\ 
	& $1.3$		  & $149$	&&&&April 1991&(4)\\
	& $1.346\pm0.009$ & $144.5\pm0.2$ &&&&25/08/1996&(5)\\
	& $1.34 \pm0.04$  & $147$	& 0.45&0.41&0.39&06-08/06/1987&(6)\\
					&&& $0.49\pm0.03$ & $0.49\pm0.04$ & $0.52\pm0.05$ &01/07/1996&(7)\\
\hline
TY CrA B-C &\multicolumn{5}{l}{eclipsing spectroscopic binary}&&(13,14)\\
A-BC	& $4.1$		  & $138$ &&&&&(15)\\
A-D	& $0.294\pm0.007$ & $188.5\pm1$ &&&&31/03/2002&(8)\\
\hline
R CrA	&\multicolumn{7}{l}{spectro-astrometric binary according to Takami et al. 2003,
		but no binarity detected in Bailey 1998}\\
\hline
T CrA	& $\ge 0.076\pm0.005$ & $273.0\pm1.4$ &&&& 26/08/1996 & (5)\\
	& $\ge 0.140\pm0.009$ & $277.7\pm1.3$ &&&& 26/06/1997 & (5)\\
\hline
VV CrA	& $2.1\pm0.1$	& $48\pm2$	&&& $0.96\pm0.02$ (b)&12/04/1995&(2)\\ 
	& $2.08\pm0.025$& $46.5\pm0.12$ &&&&1993/1994&(9)\\ 
	& $1.9$		& $44$		&&&&April 1991&(4)\\
	& $2.10$	& $44$		& 0.98& 0.26& 0.12&06-08/06/1987&(6)\\ 
					&&& $0.1\pm0.02$ & $0.3\pm0.03$ & $0.95\pm0.06$ &01/07/1996&(7)\\
\hline
CrAPMS 3 & $4.5$  & 58	&&&&April 1991&(4)\\
	 & $4.66$ & 58	&0.12&0.13&0.15&06-08/06/1987&(6)\\
			&&&&&0.19&&(10)\\
			&&& $0.14\pm0.01$ & $0.15\pm0.01$ & $0.17\pm0.02$ &01/07/1996&(7)\\
\hline

CrAPMS 6 & $3.71$ & $207.26$ &0.95&0.95&0.89&1997&(10)\\ \hline


RXJ1857.5-3732 & $\le 4$ &&&&&&(11)\\
\hline
DENIS-PJ	& $0.065\pm0.001$~ & $279.2\pm0.1$ & &&&24/09/2002&(12)\\ 
185950-370632.9	& $0.057\pm0.0005$ & $279.1\pm0.1$ & &&&24/09/2002&(12)\\ 
		& $0.066\pm0.003$~ & $283.8\pm1.2$ & &&&12/09/2000&(12)\\ 
		& $0.059\pm0.003$~ & $271.8\pm1.2$ & &&&12/09/2000&(12)\\ 
\hline
HR 7169 & \multicolumn{7}{l}{spectroscopic binary \citep[WDS,][]{wds}, 13 arc sec CPM pair with HR 7170} \\
\hline
HR 7170 & \multicolumn{7}{l}{spectroscopic binary \citep[WDS,][]{wds}, 13 arc sec CPM pair with HR 7169} \\
\hline
HD 176386 & $\sim 4$ & $\sim 137$ & & & & & (16) \\ 

\noalign{\vskip1pt\hrule}

\end{tabular}
\end{center}


Remarks: (a) Always $\le 1$ here by definition.
(b) For consistency, we give here the reciprocal value of the flux ratio given by Ghez et al.
(c) RXJ1846.7-3636 was listed as $\sim 8''$ binary in (11), i.e. with separation above
the 6'' upper limit used here, hence regarded as two separate stars here.
(d) The very low-mass star or brown dwarf CrA 444 may also be a binary \citep{lop2005}.

References:
 (1) \citet{joy44},
 (2) \citet{Ghez97},
 (3) \citet{Baier85},
 (4) \citet{reip93},
 (5) \citet{Bailey98},
 (6) \citet{chelli95},
 (7) \citet{Prato03},
 (8) \citet{chau03},
 (9) \citet{Ageorges97},
(10) \citet{Walter97},
(11) \citet{Neuh2000},
(12) \citet{bouy2004},
(13) \citet{gap34},
(14) \citet{cas98},
(15) \citet{proust81}
(16) \citet{CCDM}
\end{table*}

Therefore, we performed this new homogeneous multiplicity survey of
most CrA members with the infrared speckle camera SHARP~I and the
infrared adaptive optics (AO) instrument ADONIS.
We explain our observations
and the data reduction in Sect. 2, list all results in Sect. 3,
compare our observations of multiples with previous observations
in Sect. 4, and conclude in Sect. 5.


\section{Observations and data reduction}

We have observed the sample of 49 optically visible young members of
the R CrA association listed in Table \ref{ObjTab}.  This list is
compiled from \citet{gp75}, \citet{mr81}, \citet{Herbig88},
\citet{Walter97}, as well as \citet{Neuh1997} and \citet{Neuh2000}.

\begin{table*}[ht]
\caption[]{Stars observed in this work}
\label{ObjTab}
%
%
%
\tabcolsep.7em
\def\MRHa{$\rm[MR81]$ H$\alpha$}
\def\GP{$\rm[GP\,75]$ R CrA}
\begin{center}
\begin{tabular}{rlcclccccc}
\noalign{\hrule\vskip1.4pt\hrule\vskip1pt}
No. & Designation	& $\alpha_{2000}$ & $\delta_{2000}$ & Date(s) of Observation(s) & Instrum. & Filter(s) & Type (a) & SpTy & Note \\
\hline
 1~& \object{S CrA}		& 19:01:08.6 & -36:57:20 & 12 July 1995, 06 July 2001	& SHARP	 & JHK & c & K6\\
 2~& \object{TY CrA}		& 19:01:40.8 & -36:52:34 & 12 July 1995			& SHARP	 & K   & H & B9\\
 3~& \object{R CrA}		& 19:01:53.6 & -36:57:08 & 12 July 1995			& SHARP	 & K   & H & A5\\
 4~& \object{DG CrA}		& 19:01:55.2 & -37:23:41 & 12 July 1995			& SHARP	 & K   & c & K0\\
 5~& \object{T CrA}		& 19:01:58.8 & -36:57:50 & 12 July 1995			& SHARP	 & K   & H & F0\\
 6~& \object{VV CrA}		& 19:03:06.7 & -37:12:50 & 12 July 1995, 06 July 2001	& SHARP	 & JHK & c & K1\\
 7~& \object{\MRHa\ 10}		& 18:58:51.8 & -37:19:23 & 12 July 1995			& SHARP	 & K   & n & K & (b)\\
 8~& \object{\MRHa\ 6}		& 19:00:01.6 & -36:37:05 & 12 July 1995			& SHARP	 & K   & c & M1\\
 9~& \object{Kn Anon 2}		& 19:01:06.9 & -36:58:07 & 12 July 1995			& SHARP	 & K   & n & G0 & (b)\\
 10~& \object{V709 CrA}		& 19:01:34.9 & -37:00:57 & 12 July 1995			& SHARP	 & K   & w & K1\\
 11~& \object{\MRHa\ 2}		& 19:01:41.6 & -36:59:53 & 12 July 1995			& SHARP	 & K   & c & K8\\
 12~& \object{V702 CrA}		& 19:02:02.0 & -37:07:44 & 12 July 1995			& SHARP	 & K   & w & G5\\
 13~& \object{CrAPMS 3}		& 19:02:22.1 & -36:55:41 & 12 July 1995, 4 July 2001	& SHARP	 & JHK & w & K2\\
 14~& \object{\MRHa\ 14}	& 19:02:27.2 & -36:58:10 & 12 July 1995, 2 July 2001	& SHARP	 & JHK & w & M0-3\\
 15~& \object{\GP\ e2}		& 19:01:27.2 & -36:59:09 & 2 July 2001			& SHARP	 & K   & T & M3-5\\
 16~& \object{\GP\ f2}		& 19:01:09.7 & -36:47:53 & 2 July 2001			& SHARP	 & K   & w & K4\\
 17~& \object{\GP\ n}		& 19:01:47.9 & -36:59:30 & 2 July 2001			& SHARP	 & K   & c &	 \\
 18~& \object{\MRHa\ 12}	& 19:00:01.7 & -36:27:58 & 6 July 2001			& SHARP	 & K   & T & M3-5\\
 19~& \object{\MRHa\ 13}	& 19:02:00.1 & -37:02:22 & 6 July 2001			& SHARP	 & K   & T & M3-5\\
 20~& \object{\MRHa\ 15}	& 19:04:17.3 & -36:59:03 & 6 July 2001			& SHARP	 & K   & T & M3-5\\
 21~& \object{V721 CrA}		& 19:09:45.9 & -37:04:26 & 2 July 2001			& SHARP	 & K   & w & K \\
 22~& \object{\MRHa\ 17}	& 19:10:43.4 & -36:59:09 & 6 July 2001			& SHARP	 & K   & T & M3-5\\
 23~& \object{CrAPMS 4 NW}	& 18:57:17.8 & -36:42:36 & 4 June 2000			& ADONIS & K   & w & M0.5\\
 24~& \object{CrAPMS 4 SE}	& 18:57:20.7 & -36:43:00 & 4 June 2000			& ADONIS & K   & w & G5\\
 25~& \object{CrAPMS 5}		& 18:58:01.7 & -36:53:45 & 4 June 2000			& ADONIS & K   & w & K5\\
 26~& \object{CrAPMS 6}		& 18:59:14.7 & -37:11:30 & 5 July 2001			& SHARP  & K   & w & M3-4\\
 27~& \object{CrAPMS 8}		& 19:00:28.9 & -36:56:02 & 2 July 2001			& SHARP  & K   & w & M3\\
 28~& \object{CrAPMS 9}		& 19:00:39.1 & -36:48:11 & 5 July 2001			& SHARP  & K   & w & M2\\
 29~& \object{RXJ1855.1-3754}	& 18:55:12.0 & -37:53:52 & 5 June 2000			& ADONIS & K   & w & K3\\
 30~& \object{RXJ1836.6-3451}	& 18:36:39.5 & -34:51:26 & 6 June 2000			& ADONIS & K   & w & M0\\
 31~& \object{RXJ1839.0-3726}	& 18:39:05.3 & -37:26:22 & 5 June 2000			& ADONIS & K   & w & K1\\
 32~& \object{RXJ1840.8-3547}	& 18:40:53.3 & -35:46:45 & 3 July 2001			& SHARP  & K   & w & M4\\
 33~& \object{RXJ1841.8-3525}	& 18:41:48.6 & -35:25:44 & 5 June 2000			& ADONIS & K   & w & G7\\
 34~& \object{RXJ1842.9-3532}	& 18:42:58.0 & -35:32:43 & 6 June 2000			& ADONIS & K   & c & K2\\
 35~& \object{RXJ1844.3-3541}	& 18:44:21.9 & -35:41:44 & 5 June 2000			& ADONIS & K   & w & K5\\
 36~& \object{RXJ1844.5-3723}	& 18:44:31.1 & -37:23:34 & 6 June 2000			& ADONIS & K   & w & M0\\
 37~& \object{RXJ1845.5-3750}	& 18:45:34.8 & -37:50:20 & 6 July 2001			& SHARP  & JHK & w & G8\\
    &				&&			 & 5 June 2000			& ADONIS & K   &	\\
 38~& \object{RXJ1846.7-3636}	& 18:46:45.6 & -36:36:18 & 4 July 2001			& SHARP  & JHK & w/w & K6-7 & (c) \\
    &				&&			 & 5 June 2000			& ADONIS & K   &	\\
 39~& \object{RXJ1852.3-3700}	& 18:52:17.3 & -37:00:12 & 6 June 2000			& ADONIS & K   & c & K3\\
 40~& \object{RXJ1853.1-3609}	& 18:53:06.0 & -36:10:23 & 5 June 2000			& ADONIS & K   & w & K2\\
 41~& \object{RXJ1856.6-3545}	& 18:56:44.0 & -35:45:32 & 6 July 2001			& SHARP  & JHK & w & M2\\
    &				&&			 & 6 June 2000			& ADONIS & K   &	\\
 42~& \object{RXJ1857.5-3732}	& 18:57:34.1 & -37:32:32 & 2 July 2001			& SHARP  & K   & w/w & M5-6\\
 43~& \object{RXJ1901.4-3422}	& 19:01:28.7 & -34:22:36 & 5 June 2000			& ADONIS & K   & w & F7 & (d)\\
 44~& \object{RXJ1901.6-3644}	& 19:01:40.5 & -36:44:32 & 6 July 2001			& SHARP  & JHK & w & M0\\
    &				&&			 & 6 June 2000			& ADONIS & K   &	\\
 45~& \object{RXJ1917.4-3756}	& 19:17:23.8 & -37:56:50 & 3 July 2001			& SHARP  & K   & w & K2\\
    &				&&			 & 5 June 2000			& ADONIS & K   &	\\
 46~& \object{HR 7169}		& 19:01:03.3 & -37:03:39 & 1 July 2001			& SHARP  & K   & H & B9\\
 47~& \object{HR 7170}		& 19:01:04.3 & -37:03:42 & 1 July 2001			& SHARP  & K   & H & B8\\
 48~& \object{HD 176386}	& 19:01:38.9 & -36:53:27 & 1 July 2001			& SHARP  & K   & H & B9\\
 49~& \object{SAO 210888}	& 19:04:44.4 & -36:50:41 & 1 July 2001			& SHARP  & K   & H & B9.5\\
\noalign{\vskip1pt\hrule}
\end{tabular}
\end{center}
Notes:
(a) c for cTTS, w for wTTS, T for TTS (unknown whether cTTS or wTTS), H for Herbig AeBe star,
    n for non-TTS.
(b) Non-members \citep{Neuh2000}. (c) Two separate TTS.
(d) Foreground star at $65 \pm 5$ pc, not a member of the CrA association \citep{Neuh2000}.
(e) All other stars are members.

\end{table*}

The majority of our targets (Table~\ref{ObjTab}) were observed with
the speckle interferometry method during two observing runs in July
1995 and July 2001 at the European Southern Observatory (ESO) 3.5\,m
New Technology Telescope (NTT) on La Silla, Chile.  We used the
SHARP~I camera (System for High Angular Resolution Pictures) of the
Max-Planck-Institut for Extraterrestrial Physics \citep{SHARP}.  All
observations were done in the K-band at $2.2\rm\,\mu m$; some of the
stars, where a companion candidate was detected, were also observed in
the H- and J-bands at $1.2\rm\,\mu m$ and $1.6\rm\,\mu m$,
respectively.  For most targets, integration times of 0.5\,s per frame
were used.  For the bright stars R and T~CrA, the time was reduced to
0.2\,s to avoid saturating the detector.  On each target, 500 frames
were taken in 1995, or 600 frames in 2001, giving a total integration
time of 100\,s for R and T~CrA, and 250 -- 300\,s for the other
targets.  To allow background subtraction and bad pixel correction,
the telescope was moved after half of the frames had been taken, to
position the target at a different spot on the detector.

A few targets were observed in June 2000 with the AO system ADONIS and
the SHARP~II camera at the ESO 3.6\,m telescope on La Silla. We used
the K-band filter of this instrument, which has a central wavelength
of $2.177\rm\,\mu m$. The observing strategy was similar to that used
for the speckle observations, i.e.\ we took many frames with short
integration times. This allows us to use the same programs and
algorithms for data reduction. Since the AO system corrects the
atmospheric turbulence, somewhat longer integration times of 1 -- 2\,s
per frame could be used.  The targets were observed at 4 different
positions on the detector.  In total, 120 -- 240 frames were recorded
to obtain a total integration time of 240\,s for each target.  Both
SHARP~I and ADONIS/SHARP~II were used in a configuration with a
field-of-view of about $12^{\prime\prime} \times 12^{\prime\prime}$.

The detectors of the SHARP cameras use separate read-out electronics
for the four quadrants, which leads to discontinuities at the quadrant
borders.  To avoid distortions in the images of our stars, we
positioned the targets in the centers of one quadrant.  This implies
that companions at separations larger than about $3''$ might be
outside the field-of-view.  To make sure we did not miss any
companions, we searched the 2MASS point source catalog \citep{2MASS}
for sources near our targets.  We did not find any additional
companions within $6''$.

\begin{figure*}[tp]
\hbox to\hsize{%
\vbox{\hbox{\strut[MR81] H$\alpha$ 14}
\hbox to0.49\hsize{%
\includegraphics[angle=90,width=0.24\hsize]{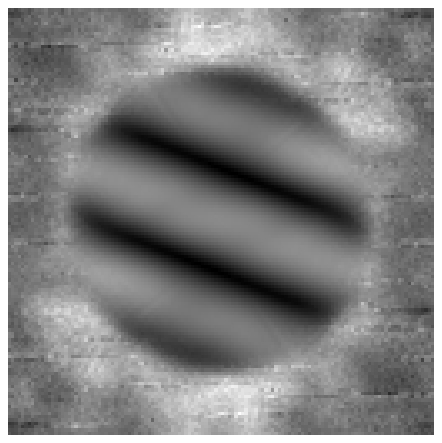}\hss	
\includegraphics[angle=90,width=0.24\hsize]{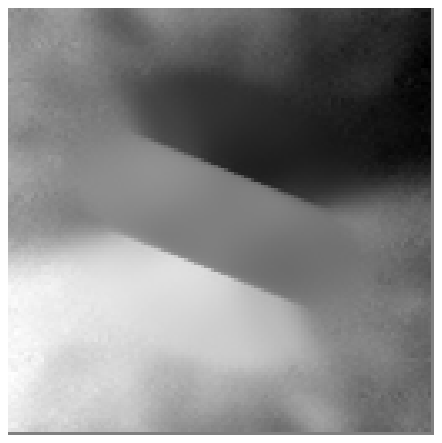}}	
}
\hfill
\vbox{\hbox{\strut[MR81] H$\alpha$ 15}%
\hbox to0.49\hsize{%
\includegraphics[angle=90,width=0.24\hsize]{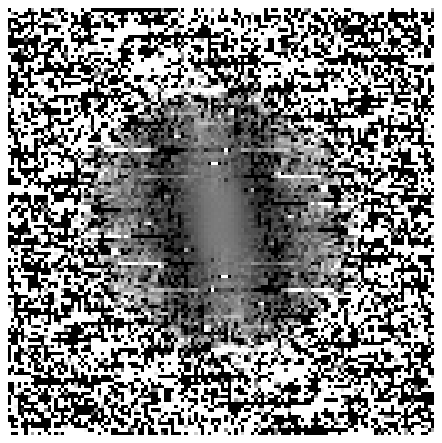}\hss	
\includegraphics[angle=90,width=0.24\hsize]{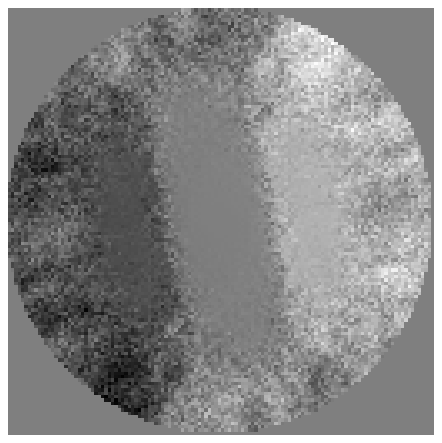}}	
}}

\medskip

\hbox to\hsize{%
\vtop{\hbox{\strut CrA PMS 8}%
\hbox to0.49\hsize{%
\includegraphics[angle=90,width=0.24\hsize]{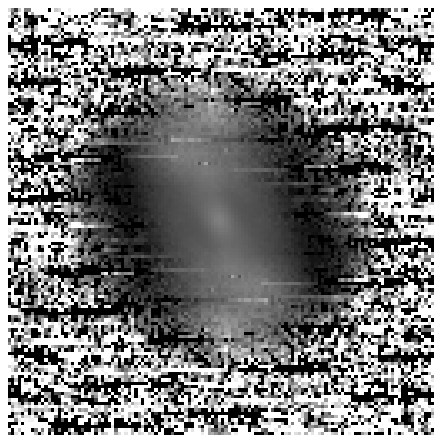}\hss	
\includegraphics[angle=90,width=0.24\hsize]{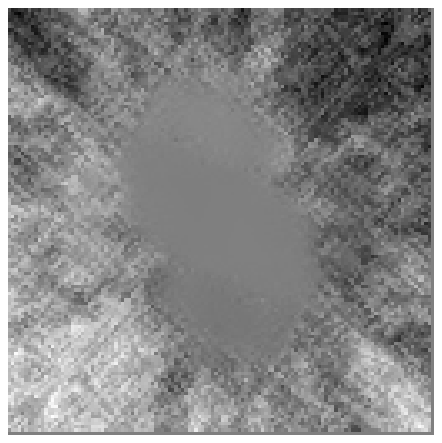}}	
}
\hfill
\vtop{\hbox{RXJ1845.5-3750}\vskip-7pt
\hbox to0.49\hsize{%
\includegraphics[angle=270,width=0.24\hsize]{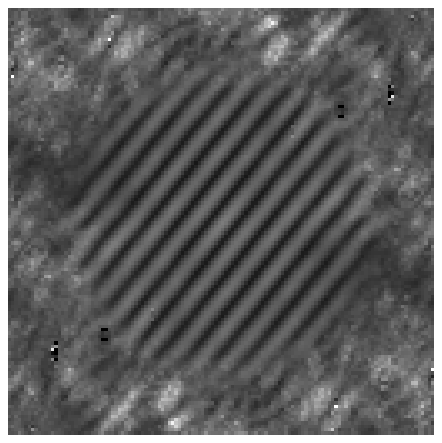}\hss	
\includegraphics[angle=270,width=0.24\hsize]{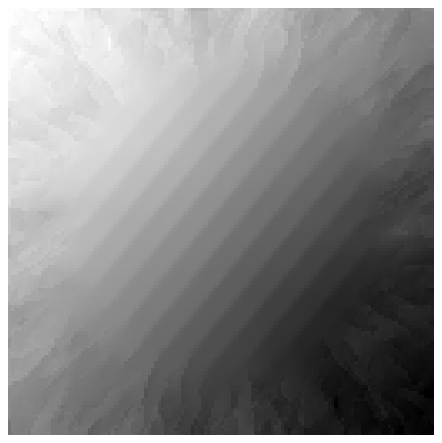}}	
}}

\medskip

\hbox to\hsize{%
\vtop{\hbox{\strut RXJ1846.7-3636 NE}
\hbox to0.49\hsize{%
\includegraphics[angle=90,width=0.24\hsize]{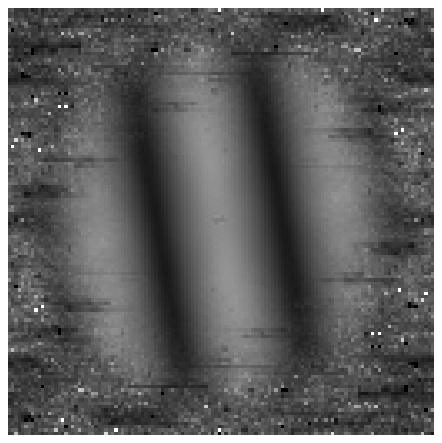}\hss	
\includegraphics[angle=90,width=0.24\hsize]{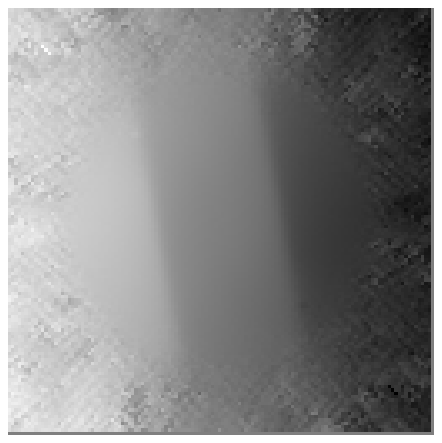}}	
}
\hfill
\vtop{\hbox{\strut RXJ1901.6-3644}
\hbox to0.49\hsize{%
\includegraphics[angle=90,width=0.24\hsize]{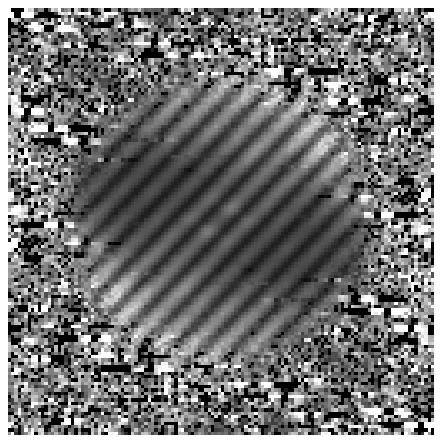}\hss	
\includegraphics[angle=90,width=0.24\hsize]{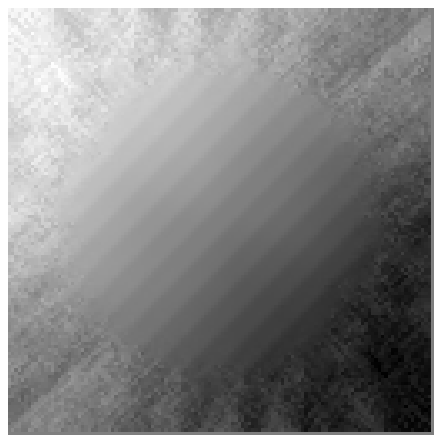}}	
}}

\medskip

\hbox to\hsize{%
\vtop{\hbox{\strut RXJ1917.4-3756}%
\hbox to0.49\hsize{%
\includegraphics[angle=90,width=0.24\hsize]{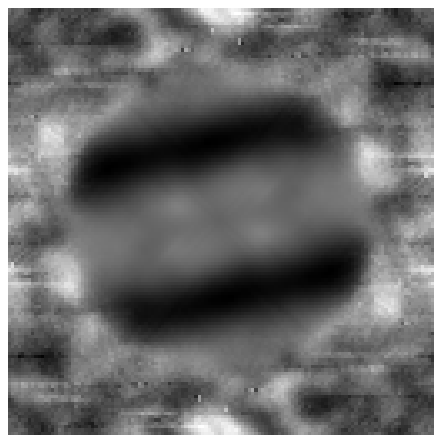}\hss	
\includegraphics[angle=90,width=0.24\hsize]{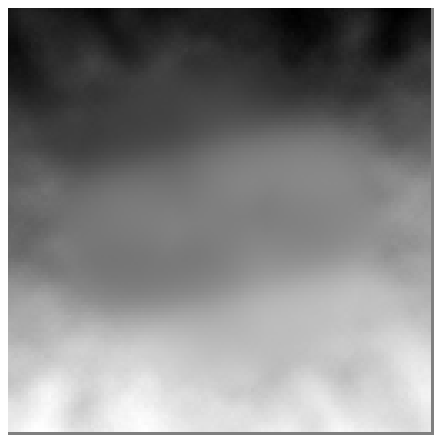}}	
}
\hfill
\vtop{\hbox{\strut HR 7170}%
\hbox to0.49\hsize{%
\includegraphics[angle=90,width=0.24\hsize]{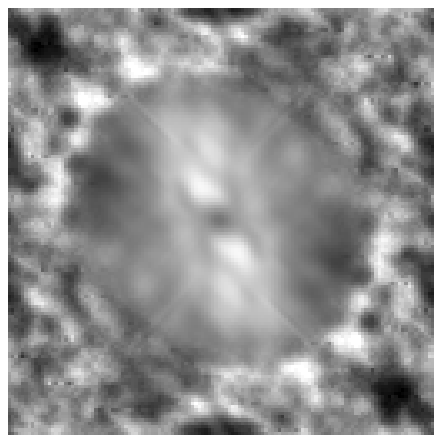}\hss	
\includegraphics[angle=90,width=0.24\hsize]{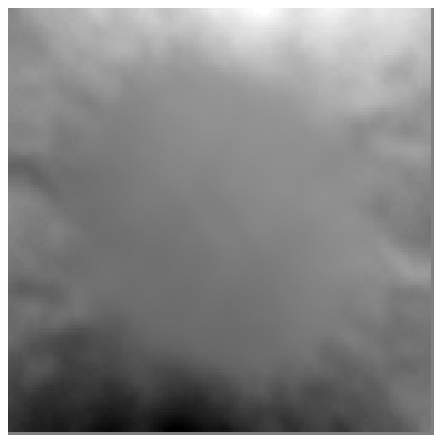}}	
}}
\caption{Complex visibilities of new binaries reconstructed from our
  SHARP~I speckle-interferometric data.
  For each binary, the modulus of the complex visibility is shown on
  the left, and the phase computed using the Knox-Thompson algorithm
  on the right.\newline
}
\label{SpeckleFig}
\end{figure*}

Although speckle interferometry can be considered by now a standard
technique \citep{Leinert92}, no program for speckle data reduction was
publicly available at the time this survey was started.  Therefore we
used our {\tt speckle} program, which was already used for binary
surveys in a number of other star-forming regions, e.g.  Taurus-Auriga
\citep{Koehler98}, Scorpius-Centaurus \citep{Koehler2000}, and
Chamaeleon \citep{Koehler2001}.  In this program, the modulus of the
complex visibility (i.e.\ the Fourier transform of the object
brightness distribution) is determined from power-spectrum analysis,
the phase is computed using the Knox-Thompson algorithm
\citep{KnoxThomp74} and from the bispectrum \citep{Lohmann83}.
Figure~\ref{SpeckleFig} shows examples of reconstructed complex
visibilities.  For a more detailed description see
\citet{Koehler2000}.

\begin{figure}[t]
\leftline{\strut CrA PMS 4NW}%
\hbox to\hsize{%
\includegraphics[angle=0,width=0.49\hsize]{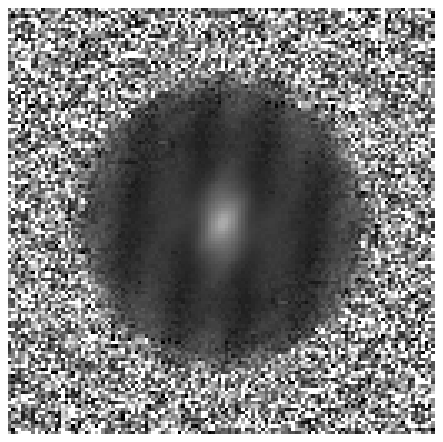}\hss	
\includegraphics[angle=0,width=0.49\hsize]{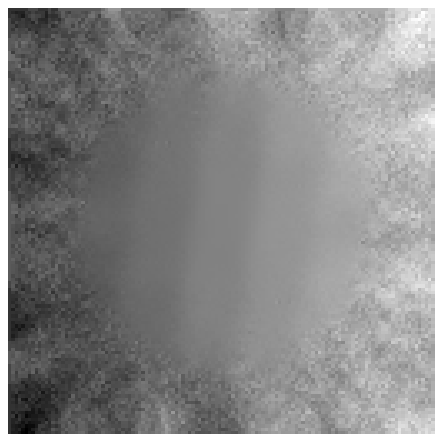}} 	

\smallskip

\leftline{\strut RXJ1844.3-3541}
\hbox to\hsize{%
\includegraphics[angle=0,width=0.49\hsize]{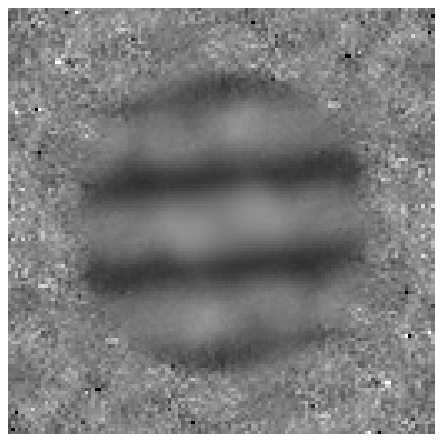}\hss	
\includegraphics[angle=0,width=0.49\hsize]{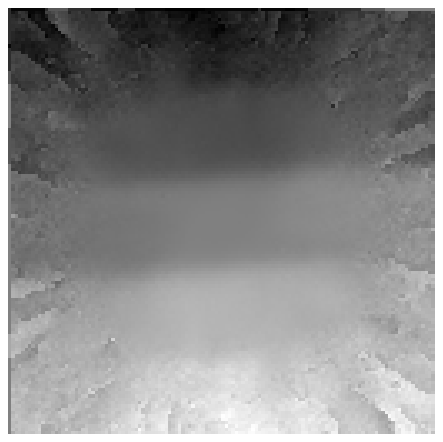}} 	

\smallskip

\leftline{\strut RXJ1853.1-3609}
\hbox to\hsize{%
\includegraphics[angle=0,width=0.49\hsize]{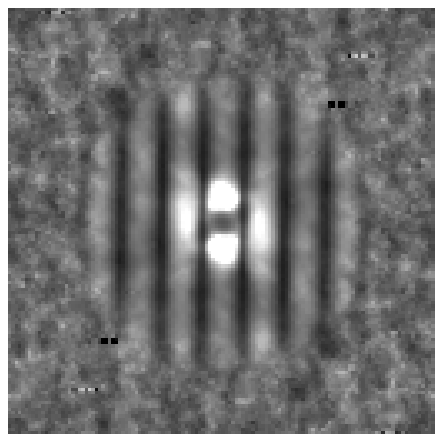}\hss	
\includegraphics[angle=0,width=0.49\hsize]{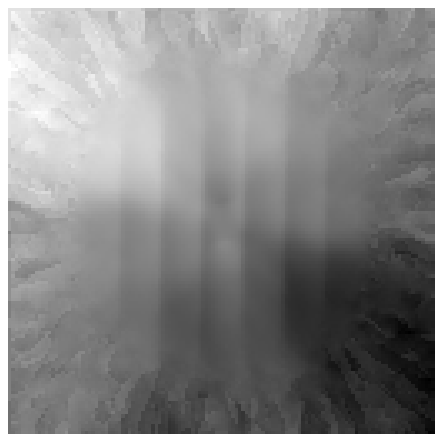}} 	
\caption{Like Fig.~\ref{SpeckleFig}, but for new binaries discovered
  in our ADONIS data.}
\label{VisBinAO}
\end{figure}


\begin{figure}[ht]
\hbox to\hsize{%
\includegraphics[width=0.5\hsize,angle=0]{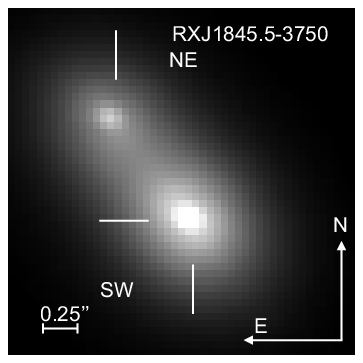}	
\includegraphics[width=0.5\hsize,angle=0]{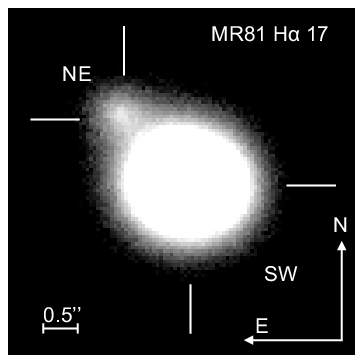}}	
\hbox to\hsize{%
\includegraphics[width=0.5\hsize,angle=0]{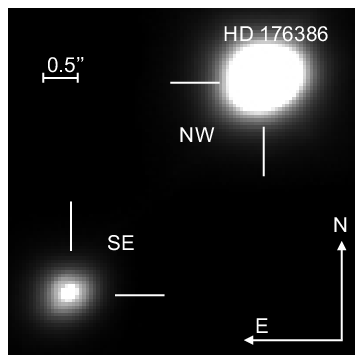} 	
\includegraphics[width=0.5\hsize,angle=0]{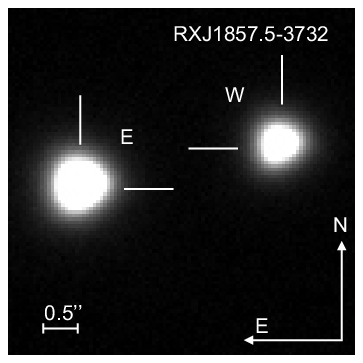}}	
\hbox to\hsize{%
\includegraphics[width=0.5\hsize,angle=0]{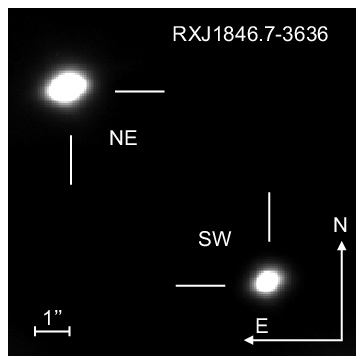} 	
\includegraphics[width=0.5\hsize,angle=0]{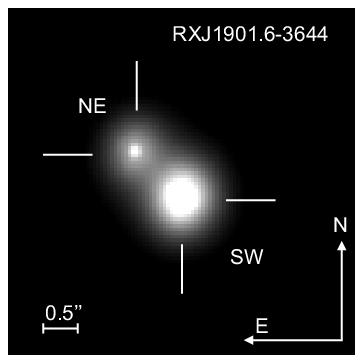}}	
\hbox to\hsize{%
\includegraphics[width=0.5\hsize,angle=0]{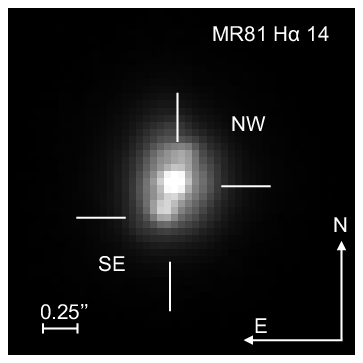}	
\includegraphics[width=0.5\hsize,angle=0]{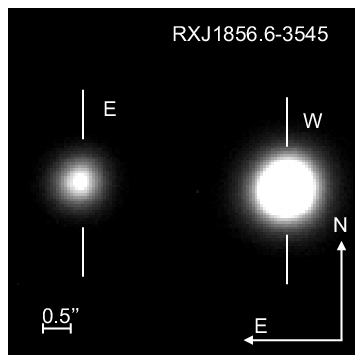}}	
\caption{Our K-band SHARP I images of wide binaries in Corona
  Australis.  RXJ~1846.7-3636 NE-SW is not regarded as binary in this
  paper, because of the 6 arc sec upper separation limit chosen;
  RXJ~1846.7-3636 NE itself is a close binary shown in
  Fig.~\ref{SpeckleFig}.}
\label{VisBinFig}
\end{figure}


If the object appeared unresolved, we computed the maximum brightness
ratio of a companion that could be hidden in the noise of the data.
The principle was to determine how far the data deviated from the
nominal result for a point source (modulus${}=1$, phase${}=0$), and
then to compute the maximum brightness ratio of a companion that would
be compatible with this amount of deviation.  This was repeated for
different position angles, and the maximum was used as upper limit for
the brightness ratio of an undetected companion. See
\citet{LeinertHenry97} for a more detailed description of this
procedure.

If the complex visibility showed the fringe pattern typical of a
binary, we computed a multidimensional least-squares fit using the
{\tt amoeba} algorithm \citep{Press92}.  Our program tried to minimize
the difference between modulus and phase computed from a model binary
and the observational data by varying the separation, position angle,
and brightness ratio of the model.  This was necessary because the
reconstructed images are a complex function of the 2-dimensional
separation vector and flux ratio that cannot be solved to compute the
binary parameters directly from the data.  Fits to different subsets
of the data yielded an estimate for the standard deviation of the
binary parameters.  We then subtracted the contribution of the
companion from the images and applied the procedure described in the
previous paragraph to find limits for the brightness of an undetected
companion.

\begin{table*}[t]
\caption[]{Companion candidates detected in this work}
\label{BinTab}
%
%
%
\tabcolsep1em
\begin{center}
\begin{tabular}{l@{~}lrc *{2}{r@{${}\pm{}$}l} l@{${}\pm{}$}lc}
\noalign{\hrule\vskip1.4pt\hrule\vskip1pt}
~Designation	&& \multicolumn{1}{c}{Date of}&	Filter &
	\omit\hfil Separation\span\omit\hfil &
	\omit\hfil Position\span\omit\hfil&
	\omit\hfil Brightness\span\omit\hfil & \omit\hfil Remarks \\
		&& \multicolumn{1}{c}{Observation} &&
	\omit\hfil [$"$]\span\omit\hfil &
	\omit\hfil Angle [$^\circ$]\span\omit\hfil&
	\omit\hfil Ratio \span\omit\hfil & \\
\hline
~S CrA		&  & 12 \,July 95~ & K & 1.346 & 0.003 & 149.3 & 0.1 & 0.432 & 0.014 & known\\
~		&  & 6 \,July 01~ & K & 1.330 & 0.004 & 150.5 & 0.2 & 0.401 & 0.016 \\
~		&  & 6 \,July 01~ & H & 1.340 & 0.009 & 150.7 & 0.3 & 0.356 & 0.012 \\
~		&  & 6 \,July 01~ & J & 1.334 & 0.003 & 150.3 & 0.2 & 0.380 & 0.011 \\ \hline
~VV CrA 	&  & 12 \,July 95~ & K & 2.082 & 0.003 & 43.5 & 0.1 & 0.515 & 0.004 & known\\
~		&  & 6 \,July 01~ & K & 2.079 & 0.007 & 43.5 & 0.2 & 0.358 & 0.013 \\
~		&  & 6 \,July 01~ & H & 2.083 & 0.006 & 43.7 & 0.2 & 0.111 & 0.008 \\
~		&  & 6 \,July 01~ & J & 2.075 & 0.015 & 43.8 & 0.2 & 0.017 & 0.001 \\ \hline
~CrAPMS 3	&  & 12 \,July 95~ & K & 4.403 & 0.018 & 57.8 & 0.2 & 0.146 & 0.003 & known\\
~		&  & 4 \,July 01~ & K & 4.442 & 0.003 & 57.5 & 0.2 & 0.151 & 0.005 \\
~		&  & 4 \,July 01~ & H & 4.452 & 0.007 & 57.5 & 0.2 & 0.125 & 0.005 \\
~		&  & 4 \,July 01~ & J & 4.455 & 0.009 & 57.4 & 0.2 & 0.106 & 0.005 \\ \hline
~ $[$MR81$]$ H$\alpha$ 14&  & 12 \,July 95~ & K & 0.229 & 0.003 & 341.0 & 0.4 & 0.877 & 0.096 & \\
~		&  & 2 \,July 01~ & K & 0.216 & 0.003 & 158.9 & 0.3 & 0.891 & 0.041 \\
~		&  & 2 \,July 01~ & H & 0.214 & 0.005 & 339.4 & 0.6 & 0.860 & 0.062 \\
~		&  & 2 \,July 01~ & J & 0.217 & 0.003 & 338.7 & 0.6 & 0.830 & 0.034 \\ \hline
~$[$MR81$]$ H$\alpha$ 15 &  & 6 \,July 01~ & K & 0.220 & 0.022 & 278.1 & 2.8 & 0.57 & 0.285 \\ \hline
~$[$MR81$]$ H$\alpha$ 17 &  & 6 \,July 01~ & K & 1.289 & 0.030 & 46.4 & 1.9 & 0.046 & 0.004 \\ \hline
~CrAPMS 4 NW	&  & 4 \,June 00~ & K & 0.361 & 0.006 & 261.7 & 0.2 & 0.178 & 0.018 \\ \hline
~CrAPMS 6	&  & 5 \,July 01~ & K & 2.363 & 0.009 & 45.4 & 0.2 & 0.835 & 0.043 & known \\ \hline
~CrAPMS 8	&  & 2 \,July 01~ & K & 0.132 & 0.009 & 110.4 & 4.3 & 0.70 & 0.300 \\ \hline
~RXJ1844.3-3541 &  & 5 \,June 00~ & K & 0.227 & 0.003 & 184.6 & 0.3 & 0.396 & 0.023 \\ \hline
~RXJ1845.5-3750 &  & 5 \,June 00~ & K & 0.904 & 0.003 & 38.8 & 0.2 & 0.602 & 0.007 \\
~		&  & 6 \,July 01~ & K & 0.886 & 0.006 & 38.6 & 0.2 & 0.542 & 0.039 \\
~		&  & 6 \,July 01~ & H & 0.898 & 0.004 & 38.5 & 0.2 & 0.568 & 0.027 \\
~		&  & 6 \,July 01~ & J & 0.891 & 0.003 & 38.5 & 0.4 & 0.508 & 0.060 \\ \hline
~RXJ1846.7-3636 & NE-SW & 5.\,June 00~ & K & 7.847 & 0.003 & 226.2 & 0.2 & 0.371 & 0.002 \\
~		& NE-SW & 4 \,July 01~ & K & 7.840 & 0.011 & 226.0 & 0.2 & 0.442 & 0.021 \\
~		& NE-SW & 4 \,July 01~ & J & 7.863 & 0.050 & 226.0 & 1.0 & 0.454 & 0.003 \\ \hline
~RXJ1846.7-3636	& NE\,AB & 5 \,June 00~ & K & 0.171 & 0.004 & 96.6 & 0.3 & 0.577 & 0.031 \\
~		& NE\,AB & 4 \,July 01~ & K & 0.171 & 0.003 & 99.9 & 0.3 & 0.680 & 0.013 \\
~		& NE\,AB & 4 \,July 01~ & J & 0.168 & 0.003 & 100.1 & 0.4 & 0.620 & 0.015 \\ \hline
~RXJ1853.1-3609 &  & 5 \,June 00~ & K & 0.516 & 0.006 & 88.7 & 0.6 & 0.518 & 0.036 \\ \hline
~RXJ1856.6-3545 &  & 6 \,June 00~ & K & 3.678 & 0.013 & 88.2 & 0.2 & 0.259 & 0.007 \\
~		&  & 6 \,July 01~ & K & 3.644 & 0.003 & 88.3 & 0.2 & 0.226 & 0.003 \\
~		&  & 6 \,July 01~ & H & 3.661 & 0.003 & 88.3 & 0.2 & 0.205 & 0.001 \\
~		&  & 6 \,July 01~ & J & 3.670 & 0.006 & 88.3 & 0.2 & 0.223 & 0.003 \\ \hline
~RXJ1857.5-3732 &  & 2 \,July 01~ & K & 2.861 & 0.003 & 281.8 & 0.2 & 0.516 & 0.011 & known \\ \hline
~RXJ1901.6-3644 &  & 6 \,June 00~ & K & 0.942 & 0.003 & 45.8 & 0.2 & 0.458 & 0.010 \\
~		&  & 6 \,July 01~ & K & 0.931 & 0.003 & 45.3 & 0.4 & 0.463 & 0.009 \\
~		&  & 6 \,July 01~ & H & 0.935 & 0.014 & 45.9 & 0.6 & 0.469 & 0.020 \\
~		&  & 6 \,July 01~ & J & 0.941 & 0.012 & 46.2 & 0.5 & 0.478 & 0.018 \\ \hline
~RXJ1917.4-3756 &  & 5 \,June 00~ & K & 0.148 & 0.005 & 198.7 & 1.4 & 0.404 & 0.035 \\
~		&  & 3 \,July 01~ & K & 0.138 & 0.003 & 194.2 & 0.9 & 0.350 & 0.024 \\ \hline
~HR 7170	&  & 1 \,July 01~ & K & 0.073 & 0.027 & 98.6 & 13.4 & 0.30 & 0.153 \\ \hline
~HD 176386	&  & 1 \,July 01~ & K & 4.104 & 0.004 & 137.4 & 0.2 & 0.128 & 0.013 & known \\
\noalign{\vskip1pt\hrule\smallskip}
\end{tabular}
\end{center}

\end{table*}

To derive the exact pixel scale and orientation of the detector, we
took images of fields in the Orion Trapezium and the Galactic center
during the observing campaigns at the NTT.  The instrumental positions
of the stars {in the Trapezium} were compared with the coordinates
given in \citet{MJMStau94} by the astrometric software
ASTROM\footnote{see
  http://www.starlink.rl.ac.uk/star/docs/sun5.htx/sun5.html}.  The
pixel scale was $49.1\pm0.1\rm\,mas/pixel$ in 1995 and
$49.5\pm0.2\rm\,mas/pixel$ in 2001.  In 1995, the detector was rotated
$(90.5\pm0.1)^\circ$ clockwise with respect to north on the sky, while
it was rotated $(90.2\pm0.2)^\circ$ counterclockwise in 2001.  We
crosschecked the results with an astrometric calibration derived
independently from images of the Galactic center (calibrated with
\citealp{Menten97}) and found that they agree within the
uncertainties.

During the ADONIS observations in June 2000, Orion could not be
observed during the night. Instead, we used images of fields near the
Galactic center and positions given by \citet{Menten97} for the
astrometric calibration.  The pixel scale was
$49.6\pm0.2\rm\,mas/pixel$, and the detector was rotated
$(179.6\pm0.1)^\circ$ clockwise.




\section{Results}

We have observed the sample of 49 stars listed in Table~\ref{ObjTab}.
Only a few known TTS members of CrA were not observed by us, namely
the stars Patten R1c, R17c, and R13a, as well as the M6-8 objects
LS-CrA 1 \& 2, Denis1859, and CrA 432 \& 444, all discovered after our
runs.  See \citet{NeuhReview2008} for a full up-to-date list of all
optically visible young members of R~CrA.

Of those 49 stars or systems, we now dismiss two ([MR 81] H$\alpha$ 10
and Kn Anon 2) from the statistical analysis, because they are
nonmembers \citep{Neuh2000}.
We also disregard RXJ1901.4-3422 from statistics, because it has
a Hipparcos-measured distance of just $65\pm5\rm\,pc$; i.e.\ it is a
foreground young star, but not a member of R CrA \citep{Neuh2000}.
(We note that neither [MR 81] H$\alpha$ 10 nor Kn Anon 2 nor
RXJ1901.4-3422 was found to be multiple.)
Then, the stars RXJ1846.7-3636 NW and SE are counted as two separate
TTS, because their separation is about 8\,arcsec, more than the
outer limit we adopt to avoid contamination by chance alignments (see
Sect.~\ref{bgsect}).  Those two objects were observed in one single
observation, both within the SHARP-I field-of-view.
The stars CrAPMS 4 NW and SE are also two separate TTS, because their
separation is about one arc minute.
We are left with 47 separately counted R CrA member systems observed,
seven Herbig Ae/Be stars, and 40 TTS.
In this sample of 47 member systems observed,
we find 19 binary stars.

\begin{figure}[ht]
\centerline{%
  \includegraphics[angle=270,width=0.9\hsize]{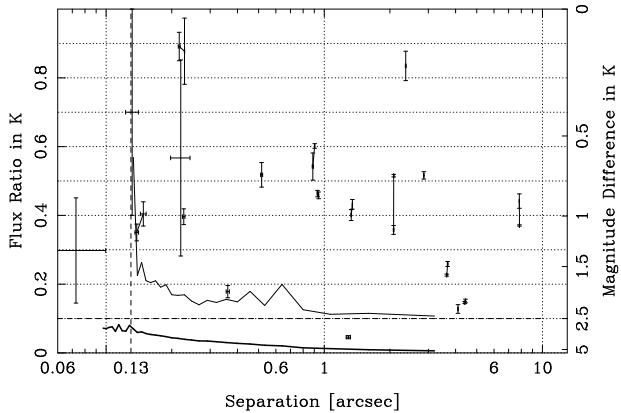}}
\caption{The results of our multiplicity survey in a plot of flux
ratio or magnitude difference vs.\ binary star separation.  The data
points mark the detected companion stars, the thick line shows the
average, and the thin line the worst sensitivity for undetected
companions.  The dashed vertical line at $0.13''$ shows the
diffraction limit for a $3.5\rm\,m$ telescope at K.  This is the limit
for unambiguous identification of binary stars.  The dashed horizontal
line shows the completeness limit in flux ratio for the whole survey.
}
\label{MaxBrgtFig}
\end{figure}

Results on separations, position angles (measured from N over E to S),
and the corresponding flux ratios in the filters used are listed
in Table \ref{BinTab} for all companions observed in this work, both
previously known companions that we detected again and newly
discovered companion candidates.
The newly discovered binaries are shown in Figs.\ \ref{SpeckleFig},
\ref{VisBinAO}, and \ref{VisBinFig}.
Figure~\ref{MaxBrgtFig} shows separation and
brightness ratio of all companions.

\subsection{Detection limits and completeness}

The modulus of the complex visibility of a binary is a cosine-shaped
function (Fig.~\ref{SpeckleFig}).  If the separation of the binary is
equal to the diffraction limit ($0.13''$ for a $3.5\rm\,m$ telescope
at K), exactly one period of the modulus of the visibility fits within
the radius where the optical transfer function of the telescope is not
zero.  Under good circumstances, it is possible to discover binaries
with even smaller separations, down to about half the diffraction
limit.\footnote{We detected a close faint companion 0.073 arc sec east
of HR 7170 (late B-type SB according to \citealp{wilk52}),
a projected separation of only 4 to 8 AU at the Hipparcos distance of
$80 \pm 20$ pc.}

However, in these cases we can only detect the first minimum, but not
the second maximum of the modulus of the visibility. Therefore, we
cannot distinguish a close binary star with certainty from an
elongated structure.  Even more important, we cannot be sure that we
find {\em all\/} companions at separations less than the diffraction
limit.  For these reasons, we limit ourselves to companions in the
separation range between $0.13''$ and $6''$.  The upper limit was
chosen so that contamination with background stars has little effect
(see Sect.~\ref{bgsect} for a detailed discussion of this problem).

Figure~\ref{MaxBrgtFig} shows not only the stars where we find
companions, but also the sensitivity of our survey, i.\,e.\ the
maximum brightness ratio of a possible undetected companion as a
function of the separation.  This sensitivity depends on factors like
the atmospheric conditions at the time of the observations and the
brightness of the target star.  We computed sensitivity limits for
each star in our survey, both stars where we find a companion and
stars where we do not find one.
Fig.~\ref{MaxBrgtFig} shows the average sensitivity limit and the
maximum of all limits, i.e.\ the envelope of the worst cases.

Only the data of a few stars are too noisy to allow detection
of all companions down to magnitude differences to the primary
of 2.5\,mag
(corresponding to a brightness ratio of 1:10).
These stars are $\rm[MR81]\,H\alpha\,13$, $\rm[MR81]\,H\alpha\,15$,
CrAPMS\,8, and RXJ1855.1-3754. Based on the number of companions
actually found, we expect less than 0.1 additional companions
above a brightness ratio of 1:10 at separations ${}>0.13''$.
Therefore, we are confident we have found all companions
with separations between $0.13''$ and $6''$ and with a magnitude
difference of less than
2.5\,mag.


\subsection{Chance alignments with background stars}
\label{bgsect}

%
%
%

We observed most of our companion candidates only on one occasion;
there is no way to tell from these data if two stars indeed form a
gravitationally bound system or if they are simply two unrelated stars
that happen to be close to each other projected onto the plane of the
sky.  To confirm companionship, we would need a 2nd-epoch image some
time later to show that both objects are co-moving; or even better,
one would have to find curvature in the orbital motion (see
e.g. \citealt{neuh2008}).  To estimate the number of chance
projections, we used the 2MASS point source catalog \citep{2MASS}.
Within the area of our survey (Right Ascension 279\degr\dots294\degr,
Declination $-41\degr\ldots-33\degr$), the catalog contains 26180
sources brighter than K=12\,mag, the sensitivity limit of our imaging
survey for close separations; 2MASS is complete down to about
K~$\simeq 14$ mag.  This corresponds to a surface density of 218
sources per square degree, or $1.7\cdot10^{-5}$ sources per square
arcsecond. Therefore, within $6''$ distance from one of our targets,
we expect $49\cdot6^2\cdot\pi\cdot1.7\cdot10^{-5} = 0.1$ field stars.
We conclude that we can safely assume that all the companions we find
are indeed gravitationally bound to their primary.


\section{Comparison with previous observations}

Among the 40 TTS observed, we detected the five known binaries again
and discovered 12 new binaries.  TY CrA BC was outside the
field-of-view of SHARP, and TY CrA D is too faint to be detected
reliably, although we see some indication of binarity in our speckle
data.
There are no known multiples among the three TTS not observed by
us. There are no known triples or higher order multiples among these
40 TTS.

We did not detect the companion to T CrA found in spectro-astrometric
observations by \citet{Bailey98} and \citet{Takami2003}.  This
companion was also not detected in the IR speckle observations of
\citet{Ghez97} and \citet{LeinertRichichi97}. The companion is only
detected through its effect in H$\alpha$, but \citet{Bailey98} reports
that the width and displacement of the H$\alpha$ line is inconsistent
with emission nebulosity or an outflow. They conclude that T~CrA is a
binary, and suggest that the companion might have strong H$\alpha$
emission, but is too faint in the K-band to be detected by speckle
interferometry.

Among the 7 Herbig Ae/Be stars observed, we resolved HR 7170 (new)
and HD 176386 (known, \citealp{CCDM}). Among all seven known Herbig
Ae/Be stars in R CrA, there is one quadruple (TY CrA as eclipsing SB
plus two resolved objects), one triple (HR 7170 as SB
plus one new resolved companion discovered in this work),
and four more binaries (HR 7169 and R CrA as SBs,
T CrA and HD 176386 as resolved binaries)\footnote{The two SBs
HR 7170 and HR 7169 form a common proper motion pair with only
13 arc sec separation, i.e.\ may form together a quintuple (including
the newly resolved visual companion to HR 7170).}.
Hence, among a total of 50 member systems (43 TTS plus 7 Herbig Ae/Be
stars), there are one quadruple, one triple, and 21 binaries,
i.e.\ 23 multiples\footnote{Among the 13 newly discovered binaries,
three are detected only by AO with ADONIS, stars not observed with
SHARP I in speckle mode; five binaries are detected with both
ADONIS and SHARP I speckle interferometry; five detected binaries
were observed only with SHARP I speckle; among the new binaries found
with SHARP I, five were detected only with speckle interferometry,
not by simple shift+add imaging.}.
{Since the triple system contains two, and the quadruple system
  three companions, there are 26 companion stars.}
In addition, among the five M6-8 type very low-mass stars and
brown dwarfs known in R CrA, there are one to two binaries;
however, we have left out the brown dwarfs and brown dwarf candidates
in our statistics, because the brown dwarf completeness is very poor
in CrA.

Given 23 multiples among 50 systems, we have a multiplicity fraction
(number of multiples divided by number of systems) of $46\pm10\,\%$.
Counting all binaries within triples and quadruples, 26 companions
in 50 systems give a companion-star frequency of $52\pm10\,\%$.
If we restrict these numbers to those systems observed in this work
and to those companions within the completeness limits of this work
(magnitude difference of lower than 2.5\,mag for separations between
0.13 and 6 arc sec), we have 17 binaries among the 47 systems observed
here, i.e. a percentage of $36\pm9\,\%$.  Since we do not find any
triples or higher order multiples, the multiplicity and companion-star
frequency are the same.

Since we have observed young CrA members both on the dark R CrA cloud
and (new ROSAT TTS) around the cloud, we can also compare the
multiplicity on-cloud with off-cloud: We have nine binaries among 30
on-cloud members ($\alpha={}$18:56 to 19:24 and $\delta=38^\circ$ to
36\degr), i.e.\ $30\pm10\,\%$, and also eight binaries among 17
off-cloud new ROSAT TTS, i.e.\ $47\pm17\,\%$. While the binaries
percentage off-cloud is higher than on-cloud, there is no
statistically significant difference.

We can now compare these numbers to other star-forming regions,
which have been observed before with a similar technique and
similar completness and sensitivity limits.

\subsection{Comparison to Main-Sequence stars}

The binary survey most commonly used for comparison is the work of
\citet[hereafter DM91]{DM91}, who studied a sample of 164 solar-type
main-sequence stars in the solar neighborhood.  They used
spectroscopic observations, complemented by direct imaging; therefore
they give the distribution of {\em periods} of the binaries.  Before
we can compare this to our results, we have to convert it into a
distribution of projected separations.  We follow the method described
in \citet{Koehler2001}.  In short, we simulate 10 million artificial
binaries with orbital elements distributed according to DM91.  Then we
compute the fraction of those binaries that could have been detected
by our observations, i.e.\ those having projected separations between
$0\farcs13$ and $6\arcsec$ at the distance of CrA (130\,pc).  The
result is that 38.4 binaries out of the DM91 sample of 164 systems
fall into this separation range, which corresponds to a companion-star
frequency of $(23.4\pm3.8)$\,\% (where the error is computed according
to binomial statistics).

\begin{figure}[ht]
\centerline{%
  \includegraphics[angle=270,width=0.9\hsize]{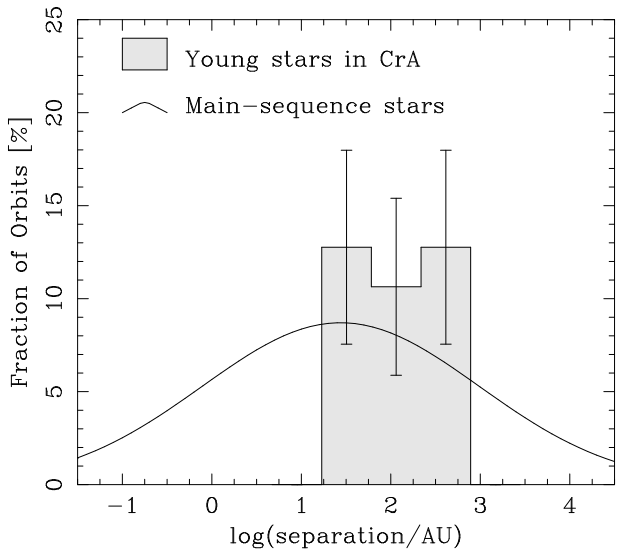}}
\caption{Companion-star frequency as a function of projected
  separation.  The histogram shows the result of this work, and the
  curve shows the distribution of separations for main-sequence
  binaries.}
\label{MSfig}
\end{figure}

Among the 47 systems in CrA, we find 17 binaries, which corresponds to
$36.2\pm8.8$ companions per 100 systems, $12.8\pm 9.6$ more than among
main-sequence stars.  Thus, the companion-star frequency in our sample
is higher by a factor of $1.5\pm 0.4$ compared to main-sequence stars.
Figure~\ref{MSfig} shows the separation distribution of our sample and
the sample of DM91.  The overabundance of binaries in CrA is somewhat
more pronounced for larger separations, which might indicate that the
peak of the separation distribution is shifted to larger separations
in CrA compared to main-sequence stars.  However, due to the
relatively small number of stars surveyed in CrA, this is not
statistically significant.

The TTS observed in our study have spectral types from G5 to M6 with
luminosity classes III and IV and with ages on the order of one to a
few Myrs, so that they have masses between $\sim 0.1$ and
1.5~M$_{\odot}$ according to \citet{Neuh2000}.  These are somewhat
lower masses than in the DM91 sample.  Unfortunately, no multiplicity
survey of M-dwarfs comparable in size to DM91 has been published.
\citet{RG97} studied the largest sample so far, which contains 81 late
K- or M-dwarfs. They find 22 companions in the separation range of our
survey in CrA, which yields a companion-star frequency of
$(27\pm6)\,\%$.  Within the errors, this is comparable to the
companion-star frequency of DM91 in the same separation range
(although the overall multiplicity found by \citet{RG97} is lower than
in DM91).  We conclude that the companion-star frequency in CrA is
also higher than among main-sequence M-dwarfs, by a factor of
$1.3\pm0.3$.

If we assume that the separation distributions of our sample in CrA
and the main-sequence sample of DM91 are the same, we can extrapolate
the number of companions to all separations.  Of DM91's 101
companions, 38.4 fall on average into the separation range surveyed by
us.  This means that the extrapolation factor is 101/38.4, which
yields a total companion-star frequency of about $(95\pm23)\,\%$.
This does {\em not} imply that 95\,\% of the stars in CrA are
multiple, but only that the average number of companions per primary
is about 0.95.  Since, e.g., triple systems contain two companions and
one primary, the fraction of multiples can be lower than 95\,\%.  We
do not know the ratio of binary to higher order multiples in CrA,
therefore we have no way to extrapolate the total number of
multiples.

\begin{table*}[htp]
\caption[]{Multiplicity in different star-forming regions}
\label{SFtab}
\begin{tabular}{lcccclcl}
\hline
Region  & $\rm S:B:T:Q$ & Multiplicity	& Companion-Star   & relative	 & Separation  & Ref. & Remarks \\
        & (a)		& $[\%]$        & Frequency $[\%]$ & to DM91	 & range ['']  &      &  \\ \hline
Tau-Aur &~$60:39:3:2$	& $42.3\pm6.4$  & $49.0\pm6.9$	   & $1.9\pm0.3$ & 0.13 -- 13  & (1) & on-cloud \\
Tau-Aur &~$24:21:0:0$	& $46.7\pm10.2$ & $46.7\pm10.2$    & $3.8\pm1.2$ & 0.09 -- 2.5 & (2) & on-cloud \\
Tau-Aur &~$40:27:2:1$	& $42.9\pm5.5$	& $48.6\pm8.3$	   & $1.9\pm0.4$ & 0.13 -- 13  & (3) & off-cloud \\
Lupus	&~$22:10:1:0$	& $33.3\pm10.1$ & $36.4\pm10.4$    & $1.4\pm0.4$ & 0.10 -- 12  & (4) & on-cloud\\
Cha	&~$19:~~9:1:0$	& $52.6\pm16.6$ & $57.9\pm17.4$	   & $2.2\pm0.7$ & 0.10 -- 12  & (4) & on-cloud\\
Cha	&~$66:11:0:0$	& $14.3\pm 4.3$ & $14.3\pm 4.3$    & $0.6\pm0.3$ & 0.13 -- 6   & (5) & off-cloud\\
Oph-Sco &~$13:11:0:0$	& $45.8\pm13.8$ & $45.8\pm13.8$	   & $2.9\pm0.9$ & 0.09 -- 2.5 & (2) & on-cloud\\
Oph	&$114:42:2:0$	& $27.8\pm4.2$	& $29.1\pm4.3$	   & $1.2\pm0.3$ & 0.13 -- 6.4 & (6) & on-cloud \\
Sco-Cen &~$59:27:2:0$	& $32.6\pm6.1$	& $35.2\pm6.3$	   & $1.6\pm0.3$ & 0.13 -- 6   & (7) & off-cloud \\
\hline
CrA     &~$30:17:0:0$	& $36.2\pm8.8$ 	& $36.2\pm8.8$ 	   & $1.5\pm0.4$ & 0.13 - 6   & (8) & total \\
CrA     &~$21:9:0:0$	& $30.0\pm10.0$	& $30.0\pm10.0$	   & $1.3\pm0.5$ & 0.13 - 6   & (8) & on-cloud \\
CrA     &~~$9:8:0:0$	& $47.1\pm16.6$	& $47.1\pm16.6$	   & $2.0\pm0.7$ & 0.13 - 6   & (8) & off-cloud \\ \hline
\end{tabular}

Remarks:
(a) number of singles (S) to binaries (B) to triples (T) to quadruples (Q).\\
References:
(1) \citet{Leinert93};
(2) \citet{ghez93};
(3) \citet{Koehler98};
(4) \citet{Ghez97};
(5) \citet{Koehler2001};
(6) \citet{Ratzka2005};
(7) \citet{Koehler2000};
(8) this work.
\end{table*}

\subsection{Comparison to other star-forming regions}

Table~\ref{SFtab} lists the results of multiplicity surveys in several
star-forming regions.  Since the separation ranges where these surveys
could detect companions are not always the same, it is not useful to
compare the companion-star frequencies directly.  Instead, it is
standard practice to divide the companion-star frequency by the
frequency of binaries among main-sequence stars (DM91) in the same
separation range, and compare the resulting relative factors.

In most T- and OB-associations, the companion-star frequency is found
to be significantly higher than among main-sequence stars, by factors
of 1.5 -- 2 or even more.  CrA shows the same high number of binaries
and multiples, with a factor of $1.5\pm0.4$.

If the separations of all the binaries found in two surveys are
published, it is possible to count the companions in the separation
range common to both surveys and to compare the numbers directly.  We
did this with our survey in CrA and the survey in Taurus-Auriga by
\citet{Leinert93} and \cite{Koehler98}.  The separation range in CrA
(0$\farcs$13 -- 6$''$) corresponds to 16.9 -- 780\,AU, or 0$\farcs$12
-- 5$\farcs$57 at the distance of Taurus-Auriga.  This range is fully
covered by the survey of \citet{Leinert93} and \citet{Koehler98}. They
find 65 companions in this separation range, resulting in a
companion-star frequency of $37.4\pm4.6\,\%$.  In CrA, we find
$36.2\pm8.8\,\%$, which is almost identical.  We conclude that the
companion-star frequency in CrA is as high as in similar star-forming
regions.


\section{Summary and conclusions}

We found 13 new binaries among young stars in the R CrA association
by simple speckle imaging, speckle interferometry, and AO imaging,
all with 3.5--3.6 meter telescopes.

The companion found next to [MR 81] H$\alpha$ 17 should be mentioned
explicitely (Fig.\ \ref{VisBinFig}):
It has a brightness ratio of only $0.046 \pm 0.004$ between the faint
NE component and the bright SW component (Table~\ref{BinTab}). The
unresolved system (i.e.\ the bright component) has a spectral type of
M3-5 \citep{pat98}. Hence, the fainter object ($\rm
K\simeq13.1\,mag$) may be a substellar companion or a so-called
infrared companion (e.g.\ with edge-on circumstellar disk) or a
background object. We continue to observe it in more detail (AO at
larger telescopes and spectroscopy).  All other companions found are
almost certainly normal TTS.


Among eight known cTTS, there are only two binaries ($25\pm18\,\%$),
but among 28 known wTTS, there are 13 binaries found ($46\pm13\,\%$).
No mixed systems were found, but for most companions and even some
primaries, its not known whether they are wTTS or cTTS.  The system
[MR 81] H$\alpha$ 17 SW-NE may be mixed.

The multiplicity in CrA is high and not significantly different from
other similar star-forming regions like Sco-Cen or Tau-Aur.
The multiplicity among new ROSAT TTS around the dark cloud is high,
just as it is in the case Tau-Aur.
There is no significant difference between on-cloud TTS and off-cloud
TTS in CrA, but a tendency to more binaries among the off-cloud TTS.

The multiplicity among the seven Herbig Ae/Be stars is high: one
quadruple, one triple, and four binaries (from SBs to 6 arcsec
separation). If going out to 13 arcsec separation, we would even have
one quintuple.  However, the sample of known Herbig Ae/Be stars in CrA
is too small to compare to other regions.
All triples and quadruples known in CrA are hierarchical.

For a comparison of the distribution of separations (or orbital
periods) found here in CrA with those of nearby solar-type,
main-sequence stars (DM91), see Fig.~\ref{MSfig}.  We can extrapolate
to the total companion-star frequency (at any separation) in CrA
being $\sim95\,\%$, also similar to Sco-Cen and Tau-Aur.


\acknowledgements
We thank the referee A.\ Brandeker for a thorough referee report.
This research has made use of the SIMBAD database, operated at CDS,
Strasbourg, France.
This publication makes use of data products from the Two Micron All
Sky Survey, which is a joint project of the University of
Massachusetts and the Infrared Processing and Analysis
Center/California Institute of Technology, funded by the National
Aeronautics and Space Administration and the National Science
Foundation.


\bibliographystyle{bibtex/aa}
\bibliography{CrA}

\begin{thebibliography}{46}
\expandafter\ifx\csname natexlab\endcsname\relax\def\natexlab#1{#1}\fi

\bibitem[{{Ageorges} {et~al.}(1997){Ageorges}, {Eckart}, {Monin}, \&
  {Menard}}]{Ageorges97}
{Ageorges}, N., {Eckart}, A., {Monin}, J.-L., \& {Menard}, F. 1997, \aap, 326,
  632

\bibitem[{Baier {et~al.}(1985)Baier, Keller, Weigelt, Bastian, \&
  Mundt}]{Baier85}
Baier, G., Keller, E., Weigelt, G., Bastian, U., \& Mundt, R. 1985, A\&A, 153,
  278

\bibitem[{Bailey(1998)}]{Bailey98}
Bailey, J. 1998, MNRAS, 301, 161

\bibitem[{{Bouy} {et~al.}(2004){Bouy}, {Brandner}, {Mart{\'{\i}}n}, {Delfosse},
  {Allard}, {Baraffe}, {Forveille}, \& {Demarco}}]{bouy2004}
{Bouy}, H., {Brandner}, W., {Mart{\'{\i}}n}, E.~L., {et~al.} 2004, \aap, 424,
  213, astro-ph/0404576

\bibitem[{Casey {et~al.}(1998)Casey, Mathieu, Vaz, Andersen, \&
  Suntzeff}]{cas98}
Casey, B.~W., Mathieu, R.~D., Vaz, L. P.~R., Andersen, J., \& Suntzeff, N.~B.
  1998, AJ, 115, 1617

\bibitem[{{Chauvin} {et~al.}(2003){Chauvin}, {Lagrange}, {Beust}, {Fusco},
  {Mouillet}, {Lacombe}, {Pujet}, {Rousset}, {Gendron}, {Conan}, {Bauduin},
  {Rouan}, {Brandner}, {Lenzen}, {Hubin}, \& {Hartung}}]{chau03}
{Chauvin}, G., {Lagrange}, A.-M., {Beust}, H., {et~al.} 2003, \aap, 406, L51,
  astro-ph/0304271

\bibitem[{Chelli {et~al.}(1995)Chelli, Cruz-Gonzalez, \& Reipurth}]{chelli95}
Chelli, A., Cruz-Gonzalez, I., \& Reipurth, B. 1995, A\&AS, 114, 135

\bibitem[{{Dommanget} \& {Nys}(2002)}]{CCDM}
{Dommanget}, J. \& {Nys}, O. 2002, VizieR Online Data Catalog, 1269, 0, (CCDM)

\bibitem[{{Duquennoy} \& {Mayor}(1991)}]{DM91}
{Duquennoy}, A. \& {Mayor}, M. 1991, \aap, 248, 485, (DM91)

\bibitem[{{Gaposchkin} \& {Greenstein}(1936)}]{gap34}
{Gaposchkin}, S. \& {Greenstein}, J.~L. 1936, Harvard College Observatory
  Bulletin, 904, 8

\bibitem[{{Ghez} {et~al.}(1997){Ghez}, {McCarthy}, {Patience}, \&
  {Beck}}]{Ghez97}
{Ghez}, A.~M., {McCarthy}, D.~W., {Patience}, J.~L., \& {Beck}, T.~L. 1997,
  \apj, 481, 378

\bibitem[{{Ghez} {et~al.}(1993){Ghez}, {Neugebauer}, \& {Matthews}}]{ghez93}
{Ghez}, A.~M., {Neugebauer}, G., \& {Matthews}, K. 1993, \aj, 106, 2005

\bibitem[{Glass \& Penston(1975)}]{gp75}
Glass, I.~S. \& Penston, M.~V. 1975, MNRAS, 172, 227

\bibitem[{{Herbig} \& {Bell}(1988)}]{Herbig88}
{Herbig}, G.~H. \& {Bell}, K.~R. 1988, {Third Catalog of Emission-Line Stars of
  the Orion Population} (Lick Observatory Bulletin No.\ 1111, Santa Cruz: Lick
  Observatory)

\bibitem[{{Hofmann} {et~al.}(1992){Hofmann}, {Blietz}, {Duhoux}, {Eckart},
  {Krabbe}, \& {Rotaciuc}}]{SHARP}
{Hofmann}, R., {Blietz}, M., {Duhoux}, P., {et~al.} 1992, in Progress in
  Telescope and Instrumentation Technologies, ESO Conference and Workshop
  Proceedings No. 42, ed. M.-H. {Ulrich} (ESO Garching), 617

\bibitem[{{James} {et~al.}(2006){James}, {Melo}, {Santos}, \&
  {Bouvier}}]{james2006}
{James}, D.~J., {Melo}, C., {Santos}, N.~C., \& {Bouvier}, J. 2006, \aap, 446,
  971, astro-ph/0510596

\bibitem[{Joy \& van Biesbroeck(1944)}]{joy44}
Joy, A.~H. \& van Biesbroeck, G. 1944, PASP, 56, 123

\bibitem[{Knox \& Thompson(1974)}]{KnoxThomp74}
Knox, K.~T. \& Thompson, B.~J. 1974, ApJ, 193, L45

\bibitem[{{K{\"o}hler}(2001)}]{Koehler2001}
{K{\"o}hler}, R. 2001, \aj, 122, 3325, astro-ph/0109103v1

\bibitem[{{K{\"o}hler} {et~al.}(2000){K{\"o}hler}, {Kunkel}, {Leinert}, \&
  {Zinnecker}}]{Koehler2000}
{K{\"o}hler}, R., {Kunkel}, M., {Leinert}, C., \& {Zinnecker}, H. 2000, \aap,
  356, 541

\bibitem[{K{\"o}hler \& Leinert(1998)}]{Koehler98}
K{\"o}hler, R. \& Leinert, C. 1998, \aap, 331, 977

\bibitem[{{Leinert}(1994)}]{Leinert92}
{Leinert}, C. 1994, in Lecture Notes in Physics, Berlin Springer Verlag, Vol.
  431, Star Formation and Techniques in Infrared and mm-Wave Astronomy, ed.
  T.~P. {Ray} \& S.~V.~W. {Beckwith}, 215--283

\bibitem[{Leinert {et~al.}(1997{\natexlab{a}})Leinert, Henry, Glindemann, \&
  McCarthy}]{LeinertHenry97}
Leinert, C., Henry, T., Glindemann, A., \& McCarthy, D.~W. 1997{\natexlab{a}},
  A\&A, 325, 159

\bibitem[{Leinert {et~al.}(1997{\natexlab{b}})Leinert, Richichi, \&
  M.}]{LeinertRichichi97}
Leinert, C., Richichi, A., \& M., H. 1997{\natexlab{b}}, A\&A, 318, 472

\bibitem[{Leinert {et~al.}(1993)Leinert, Zinnecker, Weitzel, Christou, Ridgway,
  Jameson, Haas, \& Lenzen}]{Leinert93}
Leinert, C., Zinnecker, H., Weitzel, N., {et~al.} 1993, \aap, 278, 129

\bibitem[{Lohmann {et~al.}(1983)Lohmann, Weigelt, \& Wirnitzer}]{Lohmann83}
Lohmann, A.~W., Weigelt, G., \& Wirnitzer, B. 1983, Appl. Opt.,, 22, 4028

\bibitem[{Lopez~Marti {et~al.}(2005)Lopez~Marti, Eisl{\"o}ffel, \&
  Mundt}]{lop2005}
Lopez~Marti, B., Eisl{\"o}ffel, J., \& Mundt, R. 2005, A\&A, 444, L175

\bibitem[{Marraco \& Rydgren(1981)}]{mr81}
Marraco, H.~G. \& Rydgren, A.~E. 1981, AJ, 86, 62

\bibitem[{McCaughrean \& Stauffer(1994)}]{MJMStau94}
McCaughrean, M.~J. \& Stauffer, J.~R. 1994, \aj, 108, 1382

\bibitem[{Menten {et~al.}(1997)Menten, Reid, Eckart, \& Genzel}]{Menten97}
Menten, K.~M., Reid, M.~J., Eckart, A., \& Genzel, R. 1997, ApJ, 475, l111

\bibitem[{{Neuh{\"a}user}(1997)}]{Neuh1997}
{Neuh{\"a}user}, R. 1997, Science, 276, 1363

\bibitem[{Neuh{\"a}user \& Forbrich(2008)}]{NeuhReview2008}
Neuh{\"a}user, R. \& Forbrich, J. 2008, in Handbook of Star Forming Regions,
  ed. B.~Reipurth (Astronomical Society of the Pacific), in press

\bibitem[{{Neuh{\"a}user} {et~al.}(2008){Neuh{\"a}user}, {Mugrauer},
  {Seifahrt}, {Schmidt}, \& {Vogt}}]{neuh2008}
{Neuh{\"a}user}, R., {Mugrauer}, M., {Seifahrt}, A., {Schmidt}, T.~O.~B., \&
  {Vogt}, N. 2008, \aap, 484, 281, astro-ph/0801.2287

\bibitem[{Neuh{\"a}user {et~al.}(2000)Neuh{\"a}user, Walter, Covino,
  {Alcal\'a}, Wolk, Frink, Guillout, Sterzik, \& {Comer\'on}}]{Neuh2000}
Neuh{\"a}user, R., Walter, F.~M., Covino, E., {et~al.} 2000, A\&A, 146, 323

\bibitem[{{Patten}(1998)}]{pat98}
{Patten}, B.~M. 1998, in Astronomical Society of the Pacific Conference Series,
  Vol. 154, Cool Stars, Stellar Systems, and the Sun, ed. R.~A. {Donahue} \&
  J.~A. {Bookbinder}, 1755

\bibitem[{Prato {et~al.}(2003)Prato, Greene, \& Simon}]{Prato03}
Prato, L., Greene, T.~P., \& Simon, M. 2003, ApJ, 584, 853

\bibitem[{{Press} {et~al.}(1992){Press}, {Teukolsky}, {Vetterling}, \&
  {Flannery}}]{Press92}
{Press}, W.~H., {Teukolsky}, S.~A., {Vetterling}, W.~T., \& {Flannery}, B.~P.
  1992, {Numerical recipes in C. The art of scientific computing, 2nd ed.}
  (Cambridge: University Press)

\bibitem[{{Proust} {et~al.}(1981){Proust}, {Ochsenbein}, \&
  {Pettersen}}]{proust81}
{Proust}, D., {Ochsenbein}, F., \& {Pettersen}, B.~R. 1981, \aaps, 44, 179

\bibitem[{{Ratzka} {et~al.}(2005){Ratzka}, {K{\"o}hler}, \&
  {Leinert}}]{Ratzka2005}
{Ratzka}, T., {K{\"o}hler}, R., \& {Leinert}, C. 2005, \aap, 437, 611,
  astro-ph/0504593

\bibitem[{Reid \& Gizis(1997)}]{RG97}
Reid, I.~N. \& Gizis, J.~E. 1997, AJ, 113, 2246

\bibitem[{Reipurth \& Zinnecker(1993)}]{reip93}
Reipurth, B. \& Zinnecker, H. 1993, A\&A, 278, 81

\bibitem[{{Skrutskie} {et~al.}(2006){Skrutskie}, {Cutri}, {Stiening},
  {Weinberg}, {Schneider}, {Carpenter}, {Beichman}, {Capps}, {Chester},
  {Elias}, {Huchra}, {Liebert}, {Lonsdale}, {Monet}, {Price}, {Seitzer},
  {Jarrett}, {Kirkpatrick}, {Gizis}, {Howard}, {Evans}, {Fowler}, {Fullmer},
  {Hurt}, {Light}, {Kopan}, {Marsh}, {McCallon}, {Tam}, {Van Dyk}, \&
  {Wheelock}}]{2MASS}
{Skrutskie}, M.~F., {Cutri}, R.~M., {Stiening}, R., {et~al.} 2006, \aj, 131,
  1163

\bibitem[{{Takami} {et~al.}(2003){Takami}, {Bailey}, \&
  {Chrysostomou}}]{Takami2003}
{Takami}, M., {Bailey}, J., \& {Chrysostomou}, A. 2003, \aap, 397, 675

\bibitem[{Walter {et~al.}(1997)Walter, Vrba, Wolk, Mathieu, \&
  Neuh{\"a}user}]{Walter97}
Walter, F.~M., Vrba, F.~J., Wolk, S.~J., Mathieu, R.~D., \& Neuh{\"a}user, R.
  1997, AJ, 114, 1544

\bibitem[{Wilson \& Joy(1952)}]{wilk52}
Wilson, R.~E. \& Joy, A.~H. 1952, ApJ, 115, 157

\bibitem[{{Worley} \& {Douglass}(1997)}]{wds}
{Worley}, C.~E. \& {Douglass}, G.~G. 1997, \aaps, 125, 523

\end{thebibliography}

\newpage\listofobjects

\end{document}